\documentclass[intlimits,twoside,a4paper]{article}

\usepackage{amsmath,amssymb}
\usepackage{graphicx,psfrag}
\usepackage{wrapfig}

\usepackage[T2A]{fontenc} 
\usepackage[cp1251]{inputenc} 

\usepackage[eqsecnum]{cmpj2}

\RequirePackage{color}

\issue{2015}{18}{4}{44601}
\doinumber{10.5488/CMP.18.44601}

\articletype{Rapid communication}

\title[On the discontinuity of the specific heat of the Ising model on a scale-free network]%
{On the discontinuity of the specific heat of the Ising model on a
scale-free network}
\author[M. Krasnytska \textsl{et al.}]
{M. Krasnytska\refaddr{label1,label2,label3}, B. Berche\refaddr{label2,label3},
Yu. Holovatch\refaddr{label1,label3}, R. Kenna\refaddr{label4,label3}}
\addresses{
\addr{label1} Institute for Condensed Matter Physics of the National Academy
of Sciences of Ukraine, \\
1 Svientsitskii St., 79011 Lviv, Ukraine
\addr{label2}Institut Jean Lamour, Universit\'e de Lorraine, F-54506 Vand\oe uvre les Nancy, France
\addr{label4} Applied Mathematics Research Centre, Coventry
University, Coventry CV1 5FB, United Kingdom
\addr{label3}Doctoral College for the Statistical Physics of Complex Systems,
Leipzig-Lorraine-Lviv-Coventry $({\mathbb L}^4)$}

\date{Received October 21, 2015}
\authorcopyright{M. Krasnytska, B. Berche, Yu. Holovatch, R. Kenna, 2015}

\begin{document}
\maketitle
\begin{abstract}
We consider the Ising model on an annealed scale-free network with node-degree distribution characterized by a power-law decay $P(K)\sim K^{-\lambda}$.
It is well established that the model is characterized by classical mean-field exponents for $\lambda>5$.
In this note we show that the specific-heat discontinuity $\delta c_h$  at the critical point
remains $\lambda$-dependent even for $\lambda>5$:  $\delta c_h=3(\lambda-5)(\lambda-1)/[2(\lambda-3)^2]$
and attains its mean-field value  $\delta c_h=3/2$  only in the limit $\lambda\to \infty$.
We compare this behaviour with recent measurements of the $d$ dependency of
$\delta c_h$  made for the Ising model on lattices with $d>4$ [Lundow~P.H., Markstr\"{o}m K.,
Nucl. Phys. B, 2015, \textbf{895}, 305].
\pacs 64.60.aq, 64.60.fd, 64.70.qd, 64.60.De
\end{abstract}

In the Ehrenfest classification, a second-order phase transition is
manifest by a  discontinuity of the second derivative of the free
energy at the transition temperature $T_\textrm{c}$ \cite{Landau_vol5}.
However, derivatives taken with respect to different thermodynamic
variables may demonstrate qualitatively different behaviour. For
magnetic systems, it is well known that the isothermal
susceptibility $\chi_T$ and magnetocaloric coefficient $m_T$ (a
mixed derivative of the free energy with respect to magnetic field
and temperature) are strongly diverging quantities, whereas the
specific heat $c_h$ often does not diverge at $T_\textrm{c}$. Considered in
the mean-field approximation, the first two quantities are singular
at $\tau=|T-T_\textrm{c}|/T_\textrm{c}=0$: $\chi_T\sim\tau^{-\gamma}$,
$m_T\sim\tau^{-\omega}$ with $\gamma^\textrm{mfa}=1$, $\omega^\textrm{mfa}=1/2$. However, the third quantity  displays a jump at $T_\textrm{c}$:
\begin{equation}\label{1}
\delta c_h=c_h(T\to T_\textrm{c}^-) - c_h(T\to T_\textrm{c}^+),
\end{equation}
with $\delta c_h^\textrm{mfa}=3/2$ and hence $c_h\sim\tau^{-\alpha}$
with $\alpha^\textrm{mfa}=0$.

For the Ising model in $d$ dimensions, the singularity of the specific
heat is  $d$-dependent: the famous Onsager solution
\cite{Onsager44} predicted $c_h(d=2)\sim \ln \tau$ (a weak
singularity with $\alpha=0$) while $\alpha(d=3)\simeq 0.109(4)$
\cite{Guida98} and $\alpha$ attains its mean-field value in
dimensions higher than the upper critical value, $\alpha(d>4)=0$.
Strictly at $d=4$, the scaling is affected by the logarithmic
correction \cite{Kenna12}
\begin{equation}\label{2}
c_h\sim \tau^{-\alpha^\textrm{mfa}} (\ln \tau)^{\hat{\alpha}}.
\end{equation}
Since $\alpha^\textrm{mfa}=0$ and the logarithmic correction-to-scaling
exponent $\hat{\alpha}=1/3$ is positive \cite{Kenna12}, the specific
heat of the Ising model diverges at $d=4$.

Although the critical exponents attain their mean-field values above
the upper critical dimension, this is not the case for critical
amplitudes. For $d>4$, the latter determine the value of the
specific heat discontinuity in equation (\ref{1}). As has been shown
recently \cite{Lundow15}, $\delta c_h$ for the Ising model at $d>4$
remains a $d$-dependent quantity that reaches the mean-field result
only in the limit $\delta c_h(d\to\infty)=3/2$. Inspired by this
observation, which was produced using Monte Carlo simulations for 5,
6, and  7-dimensional lattices \cite{Lundow15}, in this note we
analyze  the behaviour of the specific-heat discontinuity of the
Ising model on complex networks. Recent interest in  structures of
numerous natural and man-made systems \cite{networks,networks-2,networks-3,networks-4} lead, in
particular,  to the development of phase transition theory on
complex networks \cite{Dorogovtsev08}. Of particular interest are
scale-free networks, where the node-degree distribution is
characterized by a power-law decay:
\begin{equation}\label{3}
P(K)=c/K^{\lambda}\, .
\end{equation}
Here, $P(K)$ is the probability that the number of nearest
neighbours of a node (node degree) is $K$ and $c$ is a normalizing
constant. It appears that many real-world complex
networks (e.g., the internet, www, transportation networks, social networks of
 communication between people and many others) are scale-free
\cite{networks,networks-2,networks-3,networks-4}. In turn, studying properties of phase transitions
on scale-free networks may also explain peculiarities of processes
occurring on such networks too. To give just two examples, the analysis
of percolation phenomena on scale-free networks is directly related
to the stability of the network to random breakdowns or targeted
attacks, whereas the onset of an ordered phase (e.g., ferromagnetic
ordering in a spin model on a network) may correspond to a unanimous
opinion formation in a social network.

Here, the subject of our analysis is the Ising model on a complex
scale-free network. In particular, we  will consider the behaviour of the specific
heat on an annealed network.
This has been widely used
to analyze properties of various spin models (see e.g.,
\cite{annealed,annealed-2,annealed-3} and references therein).
For  annealed networks, the
links fluctuate on the same time scale as the spin variables
\cite{annealed,annealed-2,annealed-3}, therefore, the partition function is averaged both
with respect to  the link distribution and the Boltzmann
distribution. This is achieved by assigning to each node $i$ a hidden
variable $k_i$. In our particular case of a scale-free network, the
distribution of $k_i$ is given by (\ref{3}) too. The probability of
a link between any pair of nodes $(i,j)$ is chosen to be
proportional to the product $k_ik_j$ of $k$-variables on these
nodes. One can check that the expected node-degree value is then
${\rm{E}} [K_i] = k_i$. This choice leads to the Hamiltonian which,
in the absence of an external magnetic field, reads:
\begin{equation}\label{4}
{\cal H} =  - \frac{1}{N\langle k \rangle}\sum_{i > j}k_ik_j S_i S_j
.
\end{equation}
Here, $S_i=\pm 1$ is a spin variable, the sum spans all pairs of $N$
nodes and $\langle k \rangle=\sum_{i=1}^N k_i/N$ .

The prominent feature of (\ref{4}) is that the interaction term
attains a separable form. In turn, this allows for an exact
representation of the partition function via e.g.,
Stratonovich-Hubbard transformation, as it is usually done for the
Ising model on a complete graph \cite{Stanley71}, see
\cite{Krasnytska15a,Krasnytska15b} and references therein. It is straightforward to
get thermodynamic functions and, in particular, to arrive at the
conclusion that universal behaviour of the specific heat depends on
the node-degree distribution exponent $\lambda$
\cite{mfa,mfa-2,mfa-3,mfa-4}\footnote{The system remains ordered at any finite
temperature for $2< \lambda\leqslant 3$.}:
\begin{equation}\label{5}
\alpha=(\lambda-5)/(\lambda-3), \qquad 3<\lambda \leqslant 5;
\qquad \alpha=0, \qquad \lambda>5.
\end{equation}
The negativity of the exponent $\alpha$ in the region $3<\lambda <
5$ means that $\delta c_h=0$ there. Moreover, directly at
$\lambda=5$ the logarithmic correction-to-scaling exponent governs
the behaviour, similar as  for lattices at $d=4$, see equation (\ref{2}).
However, in contrast to the lattice case, the value of the exponent for
scale-free networks is negative: $\hat{\alpha}=-1$ \cite{mfa,mfa-2,mfa-3,mfa-4}. This
means that  $\delta c_h=0$ at $\lambda=5$ too.

Here, we are interested in the behaviour of the specific heat in
the region $\lambda>5$, where usual mean-field results for the
critical exponents hold. Keeping terms leading in $N$  for the
partition function, one can represent it  in the  form (see
\cite{Krasnytska15a,Krasnytska15b} for more details)
\begin{equation} \label{6}
Z_N (T)= \int_{-\infty}^{+\infty} \exp\left\{N\left[\frac{-\langle k \rangle
x^2}{2}\left(T-T_\textrm{c}\right)-\frac{\langle k^4 \rangle x^4}{12}\right]\right\} \rd x\, ,
\qquad \lambda>5,
\end{equation}
where $T_\textrm{c}={\langle k^2 \rangle}/{\langle k \rangle}$ and we have
omitted a prefactor which is not important for our analysis.

Using the method of steepest descent one finds points of maxima
($x_\star$) of the function under integration at $T>T_\textrm{c}$
($x_\star=0$) and $T<T_\textrm{c}$ ($x_\star=[-({3\langle k
\rangle}/{\langle k^4 \rangle})(T-T_\textrm{c})]^{1/2}\,$). The free energy reads:
\begin{eqnarray}\label{7}
f(T)=\left\{
      \begin{array}{ll}
        0, & \hbox{$T>T_\textrm{c}$}, \\
        -\frac{3\langle k \rangle^2}{4\langle k^4
\rangle}T(T-T_\textrm{c})^2, & \hbox{$T<T_\textrm{c}$}.
      \end{array}
    \right.
\end{eqnarray}
Correspondingly, for the specific heat one obtains
\begin{eqnarray}\label{8}
c_h=\left\{
      \begin{array}{ll}
        0, & \hbox{$T>T_\textrm{c}$}, \\
        -\frac{9\langle k \rangle^2}{2\langle k^4
\rangle}T^2+\frac{6\langle k \rangle^2}{2\langle k^4
\rangle}TT_\textrm{c}, & \hbox{$T<T_\textrm{c}$}.
      \end{array}
    \right.
\end{eqnarray}
The jump of the specific heat at $T_\textrm{c}$ is defined by the ratio
\begin{equation}\label{9}
\delta c_h=\frac{3\langle k^2 \rangle^2}{2\langle k^4 \rangle}.
  \end{equation}
Substituting the averages calculated with the distribution (\ref{3})
we obtain
\begin{equation}\label{10}
\delta c_h=\frac{3(\lambda-5)(\lambda-1)}{2(\lambda-3)^2},
\qquad \lambda>5\, .
  \end{equation}
In the limit of large $\lambda$ this delivers $\delta c_h=3/2$,
which coincides with the corresponding value on a complete graph.

\begin{figure}[t]
\centerline{
\includegraphics[angle=0, width=7.5cm]{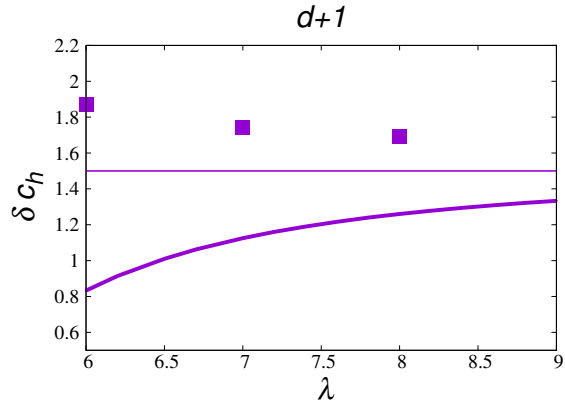}
}
\caption{
The jump in the specific heat of the Ising model on lattices at
$d>4$ (squares, results of MC simulations \cite{Lundow15}) and on an
annealed scale-free network for $\lambda>5$,  bold line equation
(\ref{10}). The thin line shows classical mean-field value $\delta
c_h=3/2$. Although $\delta c_h(\lambda\to \infty)=\delta c_h(d\to
\infty)=3/2$, the functions approach the mean-field limit from below
and from above. \label{fig1}}
\end{figure}

It is well known that Ising model on an annealed scale-free network
is characterized by classical mean-field exponents at $\lambda>5$.
As we have shown in this note, the mean-field behaviour does not
concern the specific heat jump $\delta c_h$ at $\lambda>5$. The jump
remains $\lambda$-dependent and reaches the mean-field value $\delta
c_h=3/2$ only in the limit $\lambda\to \infty$. The function $\delta
c_h(\lambda)$ is shown in figure~\ref{fig1}. Similar effect has been
observed for the Ising model on lattices at $d>4$. We show the
results of MC simulations of $d=5,6,7$-dimensional lattices
\cite{Lundow15} in the figure too. Note, that although $\delta
c_h(\lambda\to \infty)=\delta c_h(d\to \infty)=3/2$, the functions
approach the mean-field limit from below and from above.
Another
essential difference between the behaviour of $\delta c_h$  in the Ising model on
scale-free networks and on lattices  is observed directly at the
upper critical values of $\lambda$ and of $d$, respectively.
While $\alpha=0$ in both cases, the overall behaviour of $c_h$
remains singular on lattices at $d=4$ (logarithmic singularity,
$\hat{\alpha}=1/3$) whereas $\hat{\alpha}=-1$ for networks at
$\lambda=5$ and hence $\delta c_h=0$. This last case provides an
example where the logarithmic correction to scaling leads to
smoothing of behaviour of the thermodynamic function at $T_\textrm{c}$.

\section*{Acknowledgements}

This work was supported in part by FP7 EU IRSES projects  No.~295302 ``Statistical Physics in Diverse Realizations'', No.~612707 ``Dynamics of and in Complex Systems'', No.~612669
``Structure and Evolution of Complex Systems with Applications in
Physics and Life Sciences'', and by the Doctoral College for the
Statistical Physics of Complex Systems,
Leipzig-Lorraine-Lviv-Coventry $({\mathbb L}^4)$.  We thank Yuri
Kozitsky for fruitful discussions. M.K. is grateful to Klas
Markstr\"om for useful correspondence which initiated this study.

\ukrainianpart

\title{Стрибок теплоємності моделі Ізінга на безмасштабній мережі}
\author{М. Красницька\refaddr{label1,label2,label3}, Б. Берш\refaddr{label2,label3},
 Ю. Головач\refaddr{label1,label3}, Р. Кенна\refaddr{label4,label3}}
\addresses{
\addr{label1}Інститут фізики конденсованих систем НАН України, \\
вул. І.~Свєнціцького, 1, 79011 Львів, Україна
\addr{label2} Інститут Ж. Лямура,
Університет Лотарингії, F-54506 Вандувр лє Нансі, Франція
\addr{label4} Центр прикладної математики, Університет Ковентрі,
Ковентрі CV1 5FB, Англія
\addr{label3} Коледж докторантів статистичної фізики складних
систем, Ляйпціг-Лотарингія-Львів-Ковентрі $({\mathbb L}^4)$}

\makeukrtitle

\begin{abstract}
\tolerance=3000%
Ми розглядаємо модель Ізінга на відпаленій безмасштабній мережі зі
степенево-спадною функцією розподілу вузлів $P(K)\sim K^{-\lambda}$.
Відомо, що ця модель описується класичними критичними показниками
середнього поля при $\lambda>5$. Тут ми покажемо, що стрибок
теплоємності $\delta c_h$ при критичній температурі залишається
$\lambda$-залежним навіть для $\lambda>5$:  $\delta
c_h=3(\lambda-5)(\lambda-1)/[2(\lambda-3)^2]$ і досягає свого
середньопольового значення $\delta c_h=3/2$ тільки в границі $\lambda\to
\infty$. Ми порівнюємо цю поведінку із недавніми результатами
залежності $\delta c_h$ від $d$ для моделі Ізінга на гратках з $d>4$
[Lundow~P.H., Markstr\"{o}m~K., Nucl. Phys. B, 2015, \textbf{895}, 305].
\keywords  модель Ізінга, безмасштабна мережа, відпалена мережа
\end{abstract}

\end{document}